\documentclass[conference]{IEEEtran}
\IEEEoverridecommandlockouts
\usepackage{cite}
\usepackage{amsmath,amssymb,amsfonts}
\usepackage{algorithmic}
\usepackage{graphicx}
\usepackage{textcomp}
\usepackage{xcolor}
\usepackage{array} 
\usepackage{lipsum}

\def\BibTeX{{\rm B\kern-.05em{\sc i\kern-.025em b}\kern-.08em
    T\kern-.1667em\lower.7ex\hbox{E}\kern-.125emX}}
\begin{document}

\title{Carelessness Detection using Performance Factor Analysis: A New Operationalization with Unexpectedly Different Relationship to Learning\\

}

\author{\IEEEauthorblockN{Jiayi Zhang*}
\IEEEauthorblockA{\textit{Graduate School of Education} \\
\textit{University of Pennsylvania}\\
Philadelphia, PA, United States \\
joycez@upenn.edu}
~\\
\and
\IEEEauthorblockN{Ryan S. Baker}
\IEEEauthorblockA{\textit{Graduate School of Education} \\
\textit{University of Pennsylvania}\\
Philadelphia, PA, United States \\
ryanshaunbaker@gmail.com}

~\\
\and
\IEEEauthorblockN{Namrata Srivastava}
\IEEEauthorblockA{\textit{School of Engineering} \\
\textit{University of Vanderbilt}\\
Nashville, TN, United States \\
namrata.srivastava@vanderbilt.edu}
~\\
\and
\IEEEauthorblockN{Jaclyn Ocumpaugh}
\IEEEauthorblockA{\textit{Graduate School of Education} \\
\textit{University of Pennsylvania}\\
Philadelphia, PA, United States \\
jlocumpaugh@gmail.com}

\and
\IEEEauthorblockN{Caitlin Mills}
\IEEEauthorblockA{\textit{College of Education and Human Development} \\
\textit{University of Minnesota}\\
Minneapolis, MN, United States \\
cmills@umn.edu}

\and
\IEEEauthorblockN{Bruce M. McLaren}
\IEEEauthorblockA{\textit{Human-Computer Interaction Institute} \\
\textit{Carnegie Mellon University}\\
Pittsburgh, PA, United States \\
bmclaren@andrew.cmu.edu}
}

\maketitle

\begin{abstract}
Detection of carelessness in digital learning platforms has relied on the contextual slip model, which leverages conditional probability and Bayesian Knowledge Tracing (BKT) to identify careless errors, where students make mistakes despite having the knowledge. However, this model cannot effectively assess carelessness in questions tagged with multiple skills due to the use of conditional probability. This limitation narrows the scope within which the model can be applied. Thus, we propose a novel model, the Beyond-Knowledge Feature Carelessness (BKFC) model. The model detects careless errors using performance factor analysis (PFA) and behavioral features distilled from log data, controlling for knowledge when detecting carelessness. We applied the BKFC to detect carelessness in data from middle school students playing a learning game on decimal numbers and operations. We conducted analyses comparing the careless errors detected using contextual slip to the BKFC model. Unexpectedly, careless errors identified by these two approaches did not align. We found students’ post-test performance was (corresponding to past results) positively associated with the carelessness detected using the contextual slip model, while negatively associated with the carelessness detected using the BKFC model. These results highlight the complexity of carelessness and underline a broader challenge in operationalizing carelessness and careless errors.
\end{abstract}

\vspace{0.5cm}

\section{Carelessness and carelessness measurement}
\label{sec:carelessness}
Academic discussions of carelessness in classrooms date back to the 1950s \cite{eaton1956problem_behavior}. Often viewed as the result of  ineffective self-regulation, carelessness is thought to occur when students commit hurried or impulsive behaviors that result in mistakes on problems that could have been answered correctly. By distinguishing mistakes made due to carelessness from those caused by other factors, such as lack of knowledge, adaptive instruction can be provided to engage or reengage students in the effective use of self-regulation during the process of problem-solving. 

In the last several decades, two streams of work have run in parallel to investigate carelessness and detect careless behaviors. The first approach has primarily focused on self-report to identify carelessness. Grounded in a model of social problem-solving, 
D'Zurilla, Nezu \& Maydeu-Olivares \cite{dzurilla2004social_problem_solving} focused on examining the cognitive and behavioral aspects of carelessness. In their investigation, carelessness was treated as a dysfunctional problem-solving style characterized by students actively attempting to solve a problem, but making attempts that are narrow, impulsive, careless, hurried, and incomplete. Notably, from the social problem-solving model perspective, carelessness has been found to be associated with “negative” outcomes; for example, students who reported higher frequencies of carelessness were found to have lower GPAs \cite{rodriguezfornells2000impulsive}.

The other line of research on carelessness has focused on identifying carelessness from ‘careless’ behaviors. This research has primarily focused on identifying what students know and then detecting careless behavior by identifying cases where students make a mistake despite having the knowledge to answer correctly (often termed a “slip”). In the first work along this paradigm, Newman \cite{newman1977errors} and Clements \cite{clements1980errors} introduced a strategy to detect carelessness. In their studies, the same question items were administered in repeated assessments. An error was considered as careless if the student correctly solved the same item on some occasions but incorrectly on other attempts. As such, it can be inferred that the incorrectness was likely due to a slip as opposed to lack of knowledge since students demonstrated the ability to answer the same question correctly in other attempts. This general approach was extended to a broader case where carelessness is detected in different items involving the same skill \cite{baker2008bkt}. Specifically, in \cite{baker2008bkt}, a contextual slip model was proposed and then used by \cite{sanpedro2011carelessness} as an operationalization of carelessness for digital learning environments. The model estimates student knowledge and then probabilistically identifies cases where students answer a question incorrectly despite having the knowledge. As such, the erroneous answer is considered the result of carelessness as opposed to lack of knowledge. 

In a stark contrast to the results found in \cite{rodriguezfornells2000impulsive} which grounds the operationalization of carelessness in the social problem-solving model, the carelessness detected using the contextual slip model is positively associated with academic and professional achievement \cite{almeda2020stem, sanpedro2014carelessness}. Although these relationships turn negative in some cases after controlling for student knowledge \cite{sanpedro2014carelessness}, this pattern of results suggests that careless errors are more likely to be observed among otherwise successful students. 

When correlating carelessness to other behavioral and affective measures, carelessness (detected using the contextual slip model) has been found to be positively associated with engaged concentration, and negatively correlated to gaming the system, a disengaged behavior in which students attempt to succeed by exploiting properties of a learning environment \cite{fancsali2015carelessness, sanpedro2011carelessness}. Additionally, San Pedro et al. \cite{sanpedro2011carelessness} discovered a negative correlation between carelessness and confusion. Taken together, these results suggest a pattern wherein carelessness is more prevalent among high-performing and engaged students, but less common among students who are disengaged or struggling.

Even though the idea of carelessness is somewhat intuitive to grasp, there is a subtle difference in how carelessness is construed in the two streams of research, which may help explain their differential relationship with academic outcomes. In specific, there is a distinction in how student knowledge is used in determining careless errors. Knowledge plays less of a role in identifying carelessness within the social problem-solving model, whereas prior knowledge is key to identifying careless errors in the contextual slip model. This difference in how carelessness is viewed drives differences in how carelessness is measured, which likely leads to the contrasting results found in the two lines of work in terms of the relationship between carelessness and academic achievement.

\section{Limitations of the contextual slip model}
\label{sec:limitation}
In student modeling and learning analytics research, the second conceptualization has been used in several papers to detect careless errors in digital learning environments. Baker, Corbett \& Aleven’s contextual slip model \cite{baker2008bkt} uses conditional probability along with Bayesian Knowledge Tracing (BKT; \cite{corbett1995knowledge_tracing}) to estimate the probability of carelessness on a given question based on the student’s performance on the next two questions involving the same skill. In essence, it is more likely that the first incorrect answer is a careless error if the student had a high probability of already knowing the skill and then answers the next two questions containing the same skill correctly. In some cases, the carelessness estimates detected from the contextual slip model are used to train a machine-learned (ML) model which predicts carelessness from learner behavior (such as time to respond and hint use) without using future data \cite{baker2008bkt,sanpedro2011carelessness}. 

However, there are several limitations to using the contextual slip model to detect carelessness. Perhaps the most serious limitation is that this model cannot effectively assess carelessness in questions involving multiple skills, particularly those encompassing various subsets of skills. For example, consider a series of questions, where question 1 (Q1) contains skills \{A, B, C\}, Q2 contains skills \{A, C\}, Q3 contains skills \{A, D\}, and Q4 contains skills \{B, C, D\}. In this example, multiple skills are tagged in these questions, and the skills tagged in each question are not identical.

The challenge in detecting carelessness in multiple skill questions arises from two factors. First, Bayesian Knowledge Tracing (BKT) cannot easily estimate students' knowledge in questions tagged with multiple skills. BKT employs conditional probability to infer student latent knowledge on a specific skill based on historic observation on that skill \cite{corbett1995knowledge_tracing}. To apply BKT to multiple-skill items, past studies have experimented with splitting multiple-skill questions based on skills, transforming a multiple-skill question into multiple single-skill questions \cite{gong2010comparing}. Knowledge is then estimated based on each skill and aggregated across skills to predict performance on the multiple-skill item (e.g. \cite{gong2010comparing,pardos2008composition}). However, with this modification, the model is likely to overfit due to the repetition of records, where the total number of records are inflated by multiple skill items \cite{gong2010comparing}. 

A critical factor that impedes the contextual slip model’s ability to detect carelessness when items are tagged with multiple skills is its reliance on conditional probability. Specifically, the model identifies the performance of the next two questions involving the same skill and contextually estimates the probability of carelessness given the performance on these questions. However, multiple-skill items pose a challenge in identifying the next two questions with the same skill. For example, in a scenario where a student got question 1 (Q1) wrong, which contains skill \{A, B, C\}, the following questions that contain some but not all of these skills (e.g., Q2: \{A, C\}, Q3: \{A, D\}, or Q4: \{B, C, D\}) cannot provide a benchmark to contextually estimate students’ knowledge level on Q1. Because the skills only partially overlap, the student’s performance (i.e. the correctness) on the following two questions (Q2 and Q3) cannot be reliably used for the estimation of students’ knowledge level on Q1. Using the performance of Q2 and Q3 in this case would likely miscalculate the students’ knowledge level on Q1. Therefore, this approach of contextual estimation becomes infeasible for detecting carelessness in systems where items involve multiple skills.

Lastly, and acknowledged in \cite{baker2008bkt}, as the model uses the performance of the subsequent two questions, the model cannot detect carelessness in the last two questions of each skill. Therefore, the sample size may reduce significantly in detecting carelessness with this approach. \cite{baker2008bkt}’s approach to addressing this was to use the contextual slip labels to train a second contextual slip model using machine learning, that does not rely on future data; but this approach does not address the other limitations discussed above. As such, the contextual slip model has several limitations that make it less useful for many cases where carelessness detection could be of interest. 

\section{The current study}
\label{sec:study}
Given these limitations, the current paper proposes a novel model to detect carelessness in digital learning platforms. Specifically, we utilized Performance Factor Analysis (PFA; \cite{pavlik2009performance}), an alternate knowledge tracing model, to estimate students’ knowledge level. Unlike BKT, PFA can be used in cases where items are tagged with multiple skills. This PFA-based model also fits carelessness estimates step-by-step, by fitting a model that can predict carelessness from features of student actions (such as, for example, their time spent on answering a question) after controlling for student performance estimates based on past history of correctness and errors (analogous to knowledge estimation). By predicting correctness both from knowledge-related features and other features, we can take the portion of the estimate based on other features and invert it to obtain a measure of carelessness — error caused by factors other than knowledge. Using this combination of knowledge estimates and behavioral features distilled from log data, we estimate the probability of carelessness in erroneous attempts.

In contrast to the contextual slip model where carelessness is estimated primarily based on students’ knowledge (students make a mistake despite having the knowledge), the model we propose does not iteratively estimate student knowledge; instead, behavioral features are used to predict the likelihood of carelessness in an erroneous attempt. We call this model the \textbf{Beyond-Knowledge Feature Carelessness (BKFC) model}. In essence, the BKFC model uses students’ behaviors to explain the discrepancy between knowledge and performance. For example, in cases where a student is expected to know the question (i.e., high knowledge estimation) yet answers it incorrectly, the behavioral estimate is used to capture the carelessness behavior. In the current work, we use PFA to estimate knowledge; however, any knowledge tracing algorithm that has the capacity to estimate knowledge in multiple-skill items can be incorporated in the model.

We note that the BKFC model differs from the contextual slip model both conceptually and methodologically. On the conceptual front, the two models differ in what role knowledge plays in detecting careless errors. Specifically, the contextual slip model detects instances where students have knowledge but make a mistake (a “slip”). By contrast, the BKFC model detects instances where students make a mistake, but the mistake appears to be unrelated to their level of knowledge. In other words, in the BKFC model, carelessness can be detected even when students don’t know the relevant skills, as long as the specific careless behavior occurs (e.g. the impulsive, hurried, and incomplete behaviors described in the social problem-solving model). In contrast, carelessness is unlikely to be observed in the contextual slip model when students don’t have the relevant knowledge, given the model’s design. The methodology used in each model reflects this difference in conceptualization.  

In this study, we investigate the BKFC model and compare the carelessness detected across three approaches (i.e., the BKT-based contextual slip model, a machine learning model that mimics the decisions of the contextual slip model (the ML contextual slip model), and the BKFC model). To do this, we collected retrospective data from middle school students completing decimal numbers and operations items within a digital learning game, for which post-test data was also available. We applied each of these three approaches to the data. We first applied the BKT-based contextual slip model to detect carelessness in the dataset. We then generated a set of features to use within the ML contextual slip model and the BKFC model. For the ML contextual slip model, we used the identified probabilities of carelessness from the BKT-based contextual slip model as ground truth, and trained a machine learning  model to predict carelessness. For the BKFC model, we used the same features as those in the ML contextual slip model. Full mathematical details are given below. 

We then conducted a series of analyses comparing the carelessness detected across the three approaches. Specifically, we ran pairwise correlations to examine the relationship between the carelessness detected using any two of the three approaches. We also examined the relationship between carelessness detected using the three approaches and students’ post-test performance, as well as the relationship between carelessness and other disengagement and affect measures (i.e., gaming the system \cite{baker2010gaming} and 'confrustion'—when students are either confused or frustrated \cite{richey2021gaming}). Note that we originally developed the BKFC model for systems where the contextual slip model cannot be used; however, in this study, we apply BKFC to a system where the contextual slip model can also be used, for purposes of comparison and as a proof of concept.

\section{Methods}

\subsection{Learning Platforms}
\label{subsec:platforms}
We obtained retrospective data from Decimal Point, a single-player web game designed to motivate middle-school students to learn decimal concepts \cite{mclaren2017game_based_learning}. In Decimal Point, students wander through a virtual amusement park and engage in a variety of mini-games. Within each mini-game, students are first asked to solve a decimal question (problem-solving step: Figure 1.a) and then answer a multiple-choice question explaining their answer (self-explanation step: Figure 1.b). 

Five types of games are featured in the problem-solving step. Each type of game focuses on teaching (and reinforcing) one decimal skill and addressing common misconceptions with that skill.  These include (1) ordering decimals (sorting – see Figure 1a); (2) number line placement (number line); (3) decimal sequences (sequence); (4) sorting decimals into 'less-than' and 'greater-than' buckets (bucket); and (5) adding decimals (addition). 

After a student has successfully completed the problem-solving step, a self-explanation question is presented. This step is informed by prior research on the use of self-explanation, in which self-explanation has been shown to lead to deeper and more robust learning \cite{chi1994self_explanations, johnson2010self_explanation, wylie2014self_explanation}. Indeed, the self-explanations within the game have been shown to improve learning outcomes \cite{baker2024gender, mclaren2022focused}. Due to the differences in the knowledge required to answer questions in the two steps, we consider the skills in the problem-solving steps differ from the skills in their corresponding self-explanation steps. Thus, 10 skills are featured in Decimal Point in total, with 5 skills in the problem-solving steps and 5 skills in the self-explanation steps.

\begin{figure}[htbp]
\centerline{\includegraphics{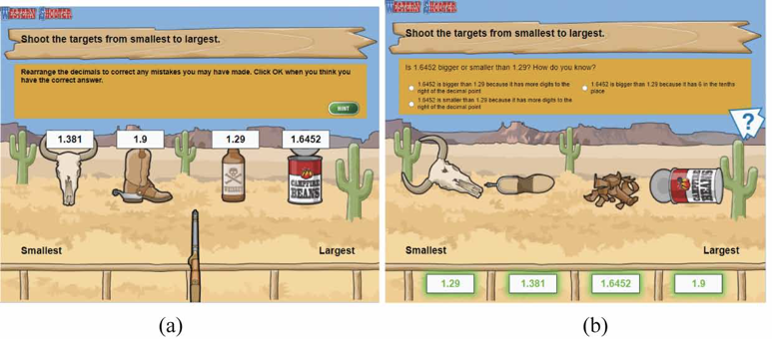}}
\caption{Problem-solving and self-explanation step in Decimal Point}
\label{fig:decimalPoint}
\end{figure}

\subsection{Data}
\label{subsec:data}
To construct models that detect carelessness, we obtained log data from Decimal Point, from a study that was conducted in 2016. In the log data, we identified the correctness of students’ initial response to each step of a problem. The binary indicator reflects whether a student gets the step correct or not on the first attempt. From here on, given the differences in skills, we consider each step as a separate question, and use the word “question” as a general term to refer to each step of a problem. 

The correctness data was used to fit BKT and PFA to estimate student knowledge level at each question (see section ~\ref{subsec:slipModel} and ~\ref{subsec:BKFC}). Log data with timestamps was used to extract features, providing additional detail around student behavior while using the system. These features were first used to train a contextual slip machine learning model to predict carelessness (see section ~\ref{subsec:MLslipModel}), and were later used in the Beyond-Knowledge Feature Carelessness (BKFC) model in estimating the probability of carelessness (discussed more in section \ref{subsec:BKFC}).

Decimal Point has only one skill per item; we chose Decimal Point in part because of this feature, which enables us to compare the novel BKFC model to the previous methods of carelessness detection. We also chose Decimal Point due to the availability of knowledge tests. We obtained data from students’ test scores on a pre-test, post-test, and delayed post-test. The post-test was administered immediately after students completed the game, and the delayed post-test was administered approximately a week after students used the program. 

Each test consisted of 24 multiple-choice questions, comprising both questions that are similar to the decimal number content presented in the game and questions that targeted underlying concepts related to decimal number operations which are not explicitly taught in the game. The performance on the knowledge tests is used to gauge the relationship between carelessness and learning. Lastly, measures of gaming the system and confrustion (when a student is either confused or frustrated) \cite{richey2021gaming}, were used to understand the relationship between carelessness and these constructs. 

In total, 181 students used Decimal Point in 2016 and had valid pre-test, post-test, and delayed post-test scores. Collectively, they answered 16,649 questions. Among them, 10,641 were answered correctly and 6,008 were answered incorrectly.

\subsection{BKT-Based Contextual Slip Model}
\label{subsec:slipModel}
Several previous papers have operationalized carelessness as contextual slip and used the BKT-based contextual slip model \cite{baker2008bkt} to detect careless errors \cite{sanpedro2011carelessness,sanpedro2014carelessness,fancsali2015carelessness,zambrano2024algorithmic}. Using the BKT-based contextual slip model, we estimate the probability of carelessness in two steps. First, we calculate the knowledge estimates of each of the 10 skills that students in our dataset practiced using the baseline BKT model. A brute-force grid search was performed to fit the parameters for the BKT models for each skill, a common approach for fitting BKT \cite{baker2010contextual}. From this baseline model, we next calculate the estimates of whether the student knew the skill at each step based on the performance on the following two questions involving the same skill using Equation \eqref{eq:contextualSlip}. The model contextually estimates the probability that an incorrect answer by a student at a specific time N is the result of slipping (or carelessness) by considering their performance on the next two questions at time N+1 and N+2 and their prior knowledge estimates. Mathematically, it is defined as a conditional probability and is directly obtainable from the probability that the student knew the skill at time N ($L_n$) given information about two subsequent actions ($A_{N+1,N+2}$).

\begin{equation}
P(A_N \text{ is a Slip} \mid A_N \text{ is incorrect}) = P(L_N \mid A_{N+1,N+2})\label{eq:contextualSlip}
\end{equation}

The underlying logic here is that if the students’ knowledge estimates for a skill was previously high and the student answers the next two questions correctly, then it is more likely that the first incorrect answer is a careless error. However, if the students’ knowledge estimates for a skill was previously low and they answer the next two questions incorrectly, then it is more likely the first incorrect answer is due to a lack of understanding. Readers are referred to \cite{baker2008bkt} for full mathematical details. Through applying the model, we obtained a probability of carelessness for each incorrect answer (except for the last two questions of each skill).

\subsection{Machine-Learned Contextual Slip Model}
\label{subsec:MLslipModel}
As previously mentioned, the BKT-based contextual slip model relies on future performance, making the detection in real-time impossible. To address this issue, \cite{baker2008bkt} use machine learning (ML) to predict carelessness in real-time using behavioral patterns. In this approach, features such as time taken and hint use are distilled to train models that predict the probability of carelessness, using machine learning algorithms. This approach successfully predicted new carelessness labels for new students \cite{baker2008bkt}.

\subsubsection{Feature Engineering}
To train models that detect carelessness, we analyzed the log data and distilled a set of features to reflect students’ behavioral patterns, drawing inspiration from features used in previous research (e.g. \cite{baker2008bkt, sanpedro2014carelessness,zambrano2024algorithmic}). As the version of the game used in this study does not provide hints, features related to hints, which have been used in previous carelessness detection papers, were not included in this study. 

In the current analysis, we extracted 17 features, which were similar to those used in previous studies and could be distilled from the existing log data. These features include: 1) the duration of a student answering a question. Using this duration, we generated features that indicate the speed of the response. Specifically, we compared the duration to the average duration of 2) other students answering the same question and 3) other students answering questions involving the same skill. A z-score was computed for each of the two comparisons. In this case, a higher z-score indicates that it took a student longer to answer a question. To examine differences in duration within individual students, we calculated 4) the difference between the duration of the current question and the student's average duration for answering all questions so far, regardless of skill. A negative value indicates that the student answered the question faster than their average duration, while a positive value indicates a slower response. Using the duration of each question, we also calculated 5) the total time a student spent answering all questions so far, 6) the total time spent answering questions involving the same skill, 7) the time spent answering the previous question, 8) the time spent answering the previous two questions, and 9) the standard deviation of the duration for the last three questions involving the current skill.

Additionally, we counted 10) the number of questions attempted so far and 11) the number of questions attempted so far involving the same skill. Three features related to students' performance were extracted. Specifically, we identified 12) whether the previous question (regardless of skill) was answered correctly, 13) the number of correct responses in the two preceding questions, regardless of skill, and 14) the percentage of previous problems involving the same skill that were answered incorrectly (percent errors).

Lastly, three features were extracted that describe how students interact with different types of questions within the learning environment. Specifically, we identified 15) the input type, which refers to the format in which students provide their response (e.g., dragging a slider on a number line, clicking a radio button, or entering text, 16) the number of actions required to complete the question, which could involve either a single action or multiple actions, and 17) the availability of gaming options, which indicates the degree to which students can exploit the feedback system to guess or adjust their answers. We define gaming options as either multiple or limited, based on the extent to which students can modify their responses using immediate feedback. For instance, when feedback is provided after each action, enabling students to make successive corrections, the gaming options are considered multiple. In contrast, when opportunities for iterative guesses or adjustments are restricted, such as in open-response formats, the gaming options are classified as limited.

\subsubsection{Training Machine Learning Model}
Using the 17 features distilled and ground-truth labeled using the BKT-based contextual slip model, we trained a machine learning model to detect carelessness. As opposed to the linear regression employed in the original study \cite{baker2008bkt}, we used random forest regressors to fit the data, as that algorithm demonstrated better performance for carelessness detection, a finding also obtained within a different data set in \cite{zambrano2024algorithmic}. The model was implemented using the Scikit-learn library \cite{pedregosa2011scikit} with default parameters. Since the labels are numerical probabilities, we used Root Mean Squared Error (RMSE) to evaluate model performance. With 5-fold student-level cross-validation, we found that the model is successful at predicting the probability of carelessness, achieving an RMSE of 0.200. This result also suggests that the features selected in the model are capable at detecting carelessness behaviors, supporting the inclusion of these features in the following BKFC model. 

\subsection{PFA-Based Beyond-Knowledge Feature Carelessness (BKFC) Model}
\label{subsec:BKFC}
As previously described, the conceptualization of carelessness has a subtle difference between the contextual slip model and the BKFC model we propose. In the BKT-based contextual slip model, carelessness is detected using conditional probability, which operationalizes carelessness as instances where students make a mistake despite having the knowledge. 

In the BKFC model we propose here, we operationalize carelessness as instances where students make a mistake, but the mistake appears to be unrelated to their level of knowledge. In other words, students may have the knowledge to answer the questions correctly, yet they made a mistake. However, within this operationalization, a careless error can occur even if the student does not have knowledge, if the same careless behavior occurs. Given this operationalization, we developed a model that detects carelessness utilizing PFA. The model uses students’ behaviors to explain the discrepancy between knowledge and performance, but distills this into a set of features that can apply regardless of whether knowledge is high or low. 

Detecting carelessness with the BKFC model involves several steps. First, we estimated the student’s knowledge level at each question using PFA; within PFA, this estimate takes the form of a prediction of the probability of student correctness. Second, we used the set of features distilled for the ML contextual slip model, which  reflect the behavioral aspect of students answering a question. For example, we distilled features such as how fast students answered a question and if the immediately previous question was answered correctly. We then fit a model using the performance estimate from step one and the behavioral features from step two to estimate the correctness of the question. This step allows us to examine how the behavioral factors influence the performance; it finds which behavioral factors are associated with performance after controlling for knowledge. We then used the behavioral features and parameters obtained from the model to estimate the probability of carelessness for incorrect questions; the initial model predicts correct answers, so we invert it to predict incorrect answers.

\subsubsection{Performance Factor Analysis}
As previously discussed, one shortcoming with the standard BKT model is its inability to estimate student knowledge for questions tagged with multiple skills. To address this issue, Pavlik et al. \cite{pavlik2009performance} proposed Performance Factors Analysis. PFA identifies the skills (also known as knowledge components; KCs) required in a question and counts the number of times a student answered these skills correctly and incorrectly in previous questions. By tracking the number of success and failures on skills in previous questions, and fitting parameters weighting each of these, PFA estimates a student’s knowledge level on the set of skills that are involved in the current question. PFA is computed using Equations \eqref{eq:pfa1} and \eqref{eq:pfa2}.

In Equation \eqref{eq:pfa1}, m is a logit value that represents the accumulated learning for student i using one or more KCs j. S counts the prior successes for the KC for the student, and f counts the prior failures for the KC for the student. The $\beta$ parameter reflects the difficulty of each KC (item difficulty variants of PFA also exist). $\gamma$ and $\rho$ are parameters that scale the effect of the success and failure counts. In the current analysis, expectation-maximization algorithm was used to fit the parameters $\beta$, $\gamma$, and $\rho$. With the parameters, we computed the value m for each question.

\begin{equation}
m(i,j \in KCs,s,f) = \sum_{j \in KCs} \left( \beta_j + \gamma_j s_{i,j} + \rho_j f_{i,j} \right)\label{eq:pfa1}
\end{equation}

With the m value computed from Equation \eqref{eq:pfa1}, Equation \eqref{eq:pfa2}, a logistic function, is then used to convert the m values to a probability. The probability indicates the likelihood of a student that has the knowledge on KCs in order to get the question correct.

\begin{equation}
p(m) = \frac{1}{1 + e^{-m}}\label{eq:pfa2}
\end{equation}

\subsubsection{Detecting Carelessness}
As previously stated, in the Beyond-Knowledge Feature Carelessness (BKFC) model, we detect carelessness by analyzing behavioral factors and examining how they account for performance, after controlling for knowledge. For instance, if a student's knowledge estimation for a question is 0.8 but they answer it incorrectly, one plausible explanation could be that the error was careless. Given this, behavioral factors that are indicative of carelessness can be used to adjust (i.e. reduce) the prediction of student performance. In this example, when taking behavioral factors into account, the predicted performance is expected to be lower.

With this conceptualization, a model was constructed to depict the relationship between students’ performance, knowledge level, and behaviors. Student performance (i.e., whether a question is answered correctly) was used as the dependent variable. Knowledge and behaviors were used as the independent variables. As shown in Equation \eqref{eq:bkfc1},  the model includes two parts – a knowledge estimation, p(m), and a behavioral function, $\omega$. The behavioral function is used to estimate how various behaviors contribute to the success of answering a question. Seventeen features are included in the behavioral function, as shown in Equation \eqref{eq:bkfc2}. We fit the model using logistic regression, for compatibility with PFA; the performance prediction from PFA and behavioral features were used jointly to estimate the probability of a student answering a question correctly.

\begin{equation}
\text{logit}(p_{\text{correct}}) = \beta_0 + \beta_1 p(m) + \omega
\label{eq:bkfc1}
\end{equation}

\begin{multline}
    \omega = \beta_2 (\text{Duration}) + 
    \beta_3 (\text{Z\_problem}) + 
    \beta_4 (\text{Z\_skill}) + \dots \\
    + \beta_{18} (\text{GamingOption})
\label{eq:bkfc2}
\end{multline}

With the fitted model, we used coefficients derived in the behavioral function to estimate carelessness (see Table \ref{tab:beta_coefficients}) in cases where students answered incorrectly. Specifically, for incorrect questions, we computed the value $\omega$, and multiplied it by -1. This was done to invert the estimation to identify careless errors. We then applied a cumulative distribution function to convert the value to a probability, given that the data is normally distributed. The probability indicates the likelihood of carelessness.

\begin{table}[htbp]
\caption{Coefficients of $\omega$}
\centering
\resizebox{\columnwidth}{!}{%
{\LARGE
\begin{tabular}{lc}
\hline
\textbf{Features} & \textbf{Unstandardized Beta Coefficients} \\
\hline
1. Duration & -0.013*** \\
2. Z-score problem & -0.032*** \\
3. Z-score problems with same skill & 0.122*** \\
4. Student difference & 0.002* \\
5. Total duration & \textless 0.001 \\
6. Total duration on the skill so far & \textless 0.001 \\
7. Duration prev question & 0.002*** \\
8. Duration prev 2 questions & 0.001 \\
9. Standard deviation of the last 3 questions & 0.003*** \\
10. Number of questions attempted & 0.001 \\
11. Number of questions attempted on the skill & 0.013*** \\
12. Previous correct & 0.006 \\
13. Previous two questions correct & 0.051*** \\
14. Percent errors & 0.528*** \\
15a. Input type\_radio button & 0.097*** \\
15b. Input type\_text & 0.099*** \\
16. Action require\_one & -0.066*** \\
17. Gaming option\_multiple & 0.199*** \\
\hline
\end{tabular}
}
}
\label{tab:beta_coefficients}
\end{table}

\subsection{Comparing Carelessness to Learning and Other Measures}
\label{subsec:compare}
After obtaining the carelessness estimates from the three approaches for the incorrect answers, we averaged the carelessness estimates at the student level and conducted a series of analyses to examine the attributes of carelessness by relating them to learning, another disengagement measure (i.e., gaming the system \cite{baker2010contextual}), and an affect measure (i.e., confrustion – when students are either confused or frustrated \cite{richey2021gaming}).

To examine the relationship between carelessness and learning, we conducted a regression analysis. In the analysis, average carelessness was used to predict students’ post-test and delayed post-test performance while controlling for the final knowledge level. The average p($L_n$) on the last question of each skill was calculated to reflect students’ final knowledge level in the regression model where carelessness was detected using the BKT-based contextual slip model or the ML contextual slip model. The average p(m) on the last question of each skill was computed to reflect the final knowledge level of a student in the regression model where carelessness was detected using the BKFC model.

Additionally, Spearman correlations were used to examine the relationship between carelessness and existing Decimal Point detectors \cite{richey2021gaming} of gaming the system and confrustion, respectively.

\section{RESULTS}

\subsection{Carelessness Detected Using the Three Approaches}
Of the 6,008 incorrect first responses on items, 4,880 could be estimated for carelessness using the BKT-based contextual slip model, as the last two questions of each skill could not be estimated \cite{baker2008bkt}. For the purpose of comparison, we only included the incorrect attempts that could be estimated by the BKT-based contextual slip model. As such, 4,880 incorrect answers were included in the following analysis, and their carelessness estimates were compared across the three approaches. 

\textbf{Carelessness Detected Using the BKT-Based Contextual Slip Model}. Using the BKT-based contextual slip model, we estimated the probability of carelessness for the 4,880 incorrect attempts. The average probability of carelessness was 0.469, with a standard deviation of 0.373. However, this average is not very meaningful, given the distribution of carelessness values shown in Figure 2.a. The bimodal distribution in Figure 2.a suggests that most of the predictions were found at the two ends of the scale.

\textbf{Carelessness Detected Using the Machine Learning Contextual Slip Model}. The average carelessness estimate detected by the ML contextual slip model is 0.393 with a standard deviation of 0.323. Since the ML model was trained to predict the ground truth (carelessness) identified by the BKT contextual slip model, the distribution of the ML carelessness estimates generally follows that of the carelessness detected by the BKT contextual slip model. Specifically, a bimodal distribution is observed; however, the difference between the two ends of the scale and the range in between was not as extreme (see Figure 2.b).

\textbf{Carelessness Detected Using the Beyond-Knowledge Feature Carelessness Model}. The average probability of carelessness in this sample was 0.473, with a standard deviation of 0.035. As shown in Figure 2.c,  the probability of carelessness detected using this approach mostly resides within the range of 0.4 and 0.6, and is normally distributed; a very different pattern than seen for the other two types of carelessness estimation. We explain this difference in the discussion section and discuss the implications of the difference between these approaches.

\begin{figure}[htp]
    \centering
    \includegraphics[width=\columnwidth]{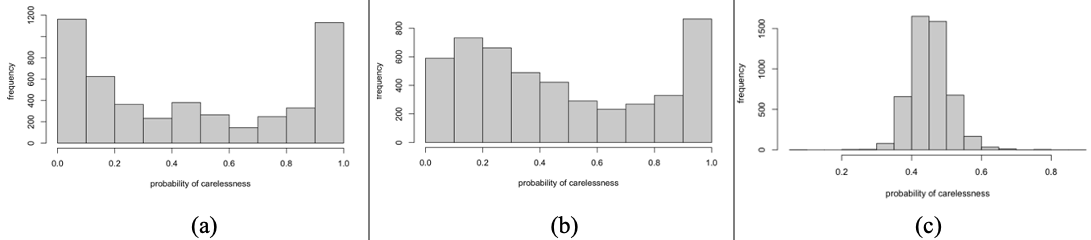} 
    \caption{Distribution of carelessness across the three models: BKT contextual slip (left), ML contextual slip (middle), BKFC (right).}
    \label{fig:single_column}
\end{figure}

\subsection{Correlation between the Three Carelessness Estimates}
Pairwise Spearman correlation was conducted to estimate the relationships between any two of the three carelessness estimates. As shown in Table \ref{tab:carelessness_correlation}, carelessness detected by the BKT-based contextual slip model and the ML contextual slip model is positively and significantly correlated. This is expected as the ML model was trained to mimic the decisions of the contextual slip model. However, carelessness detected by the BKFC model is negatively correlated with carelessness detected by either the BKT-based or ML contextual slip model.

\begin{table}[htbp]
\caption{Pairwise correlation between the three carelessness estimates}
\centering
\begin{tabular}{lcc}
\hline
\textbf{Comparison} & \textbf{$\rho$} & \textbf{$p$} \\
\hline
BKT contextual slip \~{} ML contextual slip & .849 & \textless.001 \\
BKT contextual slip \~{} BKFC              & -.21 & \textless.001 \\
ML contextual slip \~{} BKFC               & -.23 & \textless.001 \\
\hline
\end{tabular}
\label{tab:carelessness_correlation}
\end{table}
\subsection{Comparing Carelessness to Learning}
Regression analysis was conducted to examine how carelessness detected using each of the three approaches was associated with learning, controlling for the final knowledge level. Specifically, average carelessness was used to predict students’ post-test and delayed post-test scores. The average p($L_n$) or p(m) on the last question of each skill was calculated to reflect students’ final knowledge level.  

As shown in Table \ref{tab:posttest_performance}, when predicting post-test scores, we found that carelessness (however measured) and final knowledge level together explained a significant proportion of the variance in post-test scores. Specifically, the two variables explained 43\%, 44\%, and 47\% of the variance in post-test scores in models that utilized carelessness detected from the BKT contextual slip, ML contextual slip, and BKFC models, respectively. Comparing the three regression models, we found that students’ final knowledge level was positively associated with their post-test scores in all three models, indicating that the higher the final knowledge level, the higher the post-test scores. However, carelessness, as detected using the BKT-based contextual slip model or the ML contextual slip model, was positively and significantly associated with post-test scores, whereas carelessness detected using the BKFC model was negatively and significantly associated with post-test scores. Similar results were found when predicting delayed post-test scores, where carelessness detected using either the BKT-based contextual slip or the ML contextual slip model was positively and significantly associated with delayed post-test scores, while carelessness detected using the BKFC model was negatively associated with them (but not significantly).

\begin{table}[h!]
\caption{Predicting post-test and delayed post-test performance}
\resizebox{\columnwidth}{!}{%
{\LARGE
\begin{tabular}{lcc}
\hline
\textbf{Model} & \textbf{Post-test} & \textbf{Delayed post-test} \\
\hline
\textbf{BKT-based contextual slip} & $r^2 = .43, F(2, 157) = 60.43, p < .001$ & $r^2 = .47, F(2, 157) = 70.43, p < .001$ \\
Final knowledge level & $b = 16.18, t(157) = 2.64, p = .009$ & $b = 23.44, t(157) = 3.94, p < .001$ \\
Carelessness & $b = 15.43, t(157) = 3.20, p = .002$ & $b = 11.04, t(157) = 2.36, p = .02$ \\
\textbf{ML contextual slip} & $r^2 = .44, F(2, 157) = 63.59, p < .001$ & $r^2 = .47, F(2, 157) = 71.61, p < .001$ \\
Final knowledge level & $b = 17.02, t(157) = 3.23, p = .002$ & $b = 24.56, t(157) = 4.76, p < .001$ \\
Carelessness & $b = 18.28, t(157) = 3.75, p < .001$ & $b = 12.49, t(157) = 2.61, p = .01$ \\
\textbf{BKFC} & $r^2 = .47, F(2, 157) = 70.97, p < .001$ & $r^2 = .51, F(2, 157) = 83.39, p < .001$ \\
Final knowledge level & $b = 67.77, t(157) = 10.78, p < .001$ & $b = 73.33, t(157) = 12.06, p < .001$ \\
Carelessness & $b = -74.88, t(157) = -2.31, p = .02$ & $b = -48.83, t(157) = -1.56, p = .12$ \\
\hline
\end{tabular}%
}}
\label{tab:posttest_performance}
\end{table}

\subsection{Comparing Carelessness to Engagement and Affect Measures}
Using Spearman correlations, as shown in Table \ref{tab:carelessness_gaming}, we found carelessness detected using either the BKT-based contextual slip model or the ML contextual slip model is negatively associated with gaming the system. This result aligns with findings from previous studies, which found that students who made more careless errors were more engaged and were less likely to game the system \cite{fancsali2015carelessness, sanpedro2014carelessness}. 

Additionally, we found that carelessness (again detected using either the BKT-based contextual slip model or ML contextual slip model) is also significantly and negatively associated with confrustion. Similar results were found in \cite{sanpedro2014carelessness}, where students who were more careless demonstrated less confusion. However, no significant correlation was found between carelessness and frustration in that study.

However, we found carelessness detected using BKFC is positively correlated to gaming the system, the opposite of the pattern seen for the other two models. Additionally, a positive correlation is found between carelessness and confrustion, again the opposite of the pattern seen for the other models.

\begin{table}[h!]
\caption{Correlation between carelessness and other engagement and affect measures}
\resizebox{\columnwidth}{!}{%
{\LARGE
\begin{tabular}{lcc}
\hline
\textbf{Model} & \textbf{Gaming the system} & \textbf{ Confrustion} \\
\hline
BKT-based contextual slip & $r(158) = -.86, p < .001$ & $r(158) = -.87, p < .001$ \\
ML contextual slip & $r(158) = -.84, p < .001$ & $r(158) = -.86, p < .001$ \\
BKFC & $r(158) = .22, p = .005$  & $r(158) = .27, p < .001$ \\
\hline
\end{tabular}%
}}
\label{tab:carelessness_gaming}
\end{table}

\section{Discussion}
\subsection{Main Findings}
In the current work, we proposed a novel model to detect carelessness. A key benefit to this model is that it can be used in a broader set of situations compared to previous methods used to detect carelessness in digital learning platforms. The Beyond-Knowledge Feature Carelessness (BKFC) model we propose here estimates students’ knowledge level and identifies behavioral factors that cause the error other than knowledge. For example, in cases where a student is expected to know the question (i.e., high knowledge estimation) yet answers it incorrectly, the behavioral estimate is used to capture the carelessness behavior that resulted in the incorrect attempt. In this paper, we used Performance Factor Analysis to estimate students' knowledge and distilled seventeen behavioral features to capture carelessness. 

To compare this approach with previous models that detect carelessness (i.e., the BKT-based contextual slip model and the ML contextual slip model - a machine learning model that mimics the decisions of the BKT-based contextual slip model), we used all three approaches to detect careless errors in single-skill questions, where the BKT-based contextual slip model can be applied. Unexpectedly, compared to the previous approaches to detect carelessness, careless errors identified by the BKFC model had different data distributions, and differed drastically in their relationship to learning and other behavioral and affect measures. In the following paragraphs, we summarize and explain these differences and discuss the implications of these carelessness models. 

First, the estimation of carelessness differs in distribution between the three approaches. Using the BKT-based contextual slip model, we observed a bimodal distribution for carelessness estimates, such that there were two “modes” (in terms of central tendency) at the two ends of the probability scale (i.e. 0 and 1). Similar distribution was observed with carelessness estimated by the ML contextual slip model. This is expected as the ML model was trained to mimic the decision of carelessness cases identified by the BKT-based contextual slip model. By contrast, carelessness estimated using the BKFC model follows a normal distribution. This difference may be because the contextual slip model’s estimates might have been driven to extremes, either because of the conditional probability approach or because of the use of future data. On the other hand, in the BKFC model, carelessness was detected using behavioral parameters, which were derived from a logistic regression. As the behavioral parameters essentially fit to the residuals of the knowledge with performance, if these residuals were normally distributed, this may have caused the BKFC model to adopt a normal distribution. 

Second, compared to the carelessness detected using the two previous approaches, carelessness detected using the novel BKFC model differs in its relationship to learning outcomes and two other measures. We found that even when controlling for students’ final knowledge level, carelessness detected using either the BKT-based or the ML contextual slip model is positively associated with post-test and delayed post-test scores. Additionally, carelessness identified using the same approaches was found to be negatively associated with gaming the system (a disengagement measure) and negatively associated with confrustion. These results align with previous findings in \cite{fancsali2015carelessness, sanpedro2014carelessness}, showing that carelessness is more frequently observed in high-performing and more engaged students. This is possibly because, within the contextual slip approach (either the BKT-based or the ML-based), an error is only likely to be considered as carelessness after the student has acquired the skill.

In a strong contrast, carelessness detected using the BKFC model was negatively correlated with post-test and delayed post-test performance and was positively associated with gaming the system and confrustion. This suggests that the more a student games the system or becomes confused or frustrated, the more likely it is for the student to make careless errors. Similarly, a dissociable relationship was observed for performance measures, such that more careless errors were associated with worse scores on both post-test measures.

We suspect that these contrasting results in terms of the relationship between carelessness, learning outcomes, and other measures stem from a subtle difference in how carelessness is operationalized. Specifically, there is a distinction in how knowledge is leveraged to detect carelessness in these approaches. Specifically, the contextual slip model (either BKT-based or ML-based) detects instances where students make a mistake despite having the knowledge, whereas the BKFC model detects instances where students make a mistake, but the mistake appears to be unrelated to their level of knowledge. In other words, in the BKFC model, carelessness can be detected even when students don’t know the relevant skills if the same careless behavior occurs. In contrast, carelessness is unlikely to be observed in the contextual slip model when students don’t have the relevant knowledge given the model’s design.

Given the difference in the operationalization, the contextual slip model may only be able to identify carelessness within high-performing students as the model is essentially only able to detect carelessness when students have relatively high levels of knowledge. This requirement may directly lead to the positive relationship between carelessness and academic performance, as found in the current paper and previous research \cite{fancsali2015carelessness, sanpedro2011carelessness, sanpedro2014carelessness}. However, the same careless behaviors can occur when students don’t have the knowledge but make impulsive, careless, and hurried attempts, as described in the social problem-solving model \cite{dzurilla2004social_problem_solving}. The BKFC model is able to correctly capture these relationships, which presumably leads to the contrasting results seen for this method.

\subsection{Limitations and Future Work}
Despite being derived from two theoretically sound models, the carelessness detected from the contextual slip model and the novel BKFC model had very different characteristics and correlations. These contrasting results reveal a broader challenge surrounding the issue of detecting carelessness, which is deciding exactly what carelessness is and how to operationalize it. 

It is difficult to know which operationalization is correct, as there is no easy way to obtain ground truth. External observers have been unable thus far to confidently identify carelessness from screen replays, text replays, or field observations. Teachers often feel that they recognize when a student is being careless, but only when they can directly inspect step-by-step work processes to understand errors, and it is uncertain what proportion of careless errors such a procedure would capture. Self-report may provide some insight, but may also be vulnerable to demand or self-presentation effects, in unpredictable ways. There may still be value to asking students open-ended questions immediately after an incorrect attempt, to have students reflect on the reason for the error. These responses may serve as additional evidence of detecting carelessness, acting as a source of ground truth. 

In future work, we anticipate applying the BKFC model to detect carelessness in other platforms. Specifically, the primary original motivation for proposing and developing the BKFC model was to be able to detect carelessness in questions tagged with multiple skills. In these platforms, we can study this operationalization of carelessness’s relationship to other constructs, and use it within interventions that encourage students to work carefully and use appropriate self-regulated learning strategies. This operationalization’s clear negative relationship to learning outcomes makes it potentially appropriate for this type of usage. It may also be valuable to control for gaming the system and other known disengaged behaviors in future modeling work along these lines, to ensure that the most appropriate possible feedback is given.

Additionally, we hope to explore the possibility of using other knowledge tracing models in the BKFC models. Knowledge tracing models that leverage deep learning have demonstrated better performance than PFA at predicting students’ performance. Incorporating these models may potentially improve the precision of carelessness detection and make it useful in situations where PFA is not feasible (such as cases where there is not a trusted item/skill mapping).

\subsection{Conclusion}
In this paper, we proposed and described the Beyond-Knowledge Feature Carelessness (BKFC) model, a new model of carelessness that can be used when items are tagged with multiple skills. The BKFC model differs from the contextual slip model in operationalization, detecting carelessness in cases where a student makes a mistake, but the mistake appears to be unrelated to their level of knowledge. Therefore, compared to the contextual slip model, which can only detect carelessness in cases when students have the necessary knowledge, the BKFC model can detect carelessness in a broader range of cases — even when students may not have the knowledge, provided that the same careless behavior occurred. Along with the ability to detect carelessness in multiple-skill questions, the BKFC model will enable us to broaden the scope for detecting carelessness, thereby increasing its utility and capability in evaluating instances of carelessness in a broader context. Work remains to fully understand this model, but it seems to have the potential to detect carelessness in more situations and therefore to enable intervention in those situations.

\section*{Acknowledgment}
This research was supported by the National Science Foundation under grant number \#DRL-2201798. Special thanks to key members of the Decimal Point team: Aditi Haiman, Daniel Pollock, Hayden Stec, Leah Teffera, Mahboobeh Mehrvarz, and Liz Richey.

\bibliographystyle{IEEEtran}
\bibliography{main}

\begin{thebibliography}{10}
\providecommand{\url}[1]{#1}
\csname url@samestyle\endcsname
\providecommand{\newblock}{\relax}
\providecommand{\bibinfo}[2]{#2}
\providecommand{\BIBentrySTDinterwordspacing}{\spaceskip=0pt\relax}
\providecommand{\BIBentryALTinterwordstretchfactor}{4}
\providecommand{\BIBentryALTinterwordspacing}{\spaceskip=\fontdimen2\font plus
\BIBentryALTinterwordstretchfactor\fontdimen3\font minus \fontdimen4\font\relax}
\providecommand{\BIBforeignlanguage}[2]{{%
\expandafter\ifx\csname l@#1\endcsname\relax
\typeout{** WARNING: IEEEtran.bst: No hyphenation pattern has been}%
\typeout{** loaded for the language `#1'. Using the pattern for}%
\typeout{** the default language instead.}%
\else
\language=\csname l@#1\endcsname
\fi
#2}}
\providecommand{\BIBdecl}{\relax}
\BIBdecl

\bibitem{eaton1956problem_behavior}
M.~T. Eaton, L.~A. D’Amico, and B.~N. Phillips, ``Problem behavior in school,'' \emph{Journal of Educational Psychology}, vol.~47, no.~6, pp. 350--357, 1956.

\bibitem{dzurilla2004social_problem_solving}
T.~J. D’Zurilla, A.~M. Nezu, and A.~Maydeu-Olivares, ``Social problem solving: Theory and assessment,'' in \emph{Social Problem Solving: Theory, Research, and Training}, E.~C. Chang, T.~J. D’Zurilla, and L.~J. Sanna, Eds.\hskip 1em plus 0.5em minus 0.4em\relax Washington: American Psychological Association, 2004, pp. 11--27.

\bibitem{rodriguezfornells2000impulsive}
A.~Rodr\'{\i}guez-Fornells and A.~Maydeu-Olivares, ``Impulsive/careless problem-solving style as predictor of subsequent academic achievement,'' \emph{Personality and Individual Differences}, vol.~28, no.~4, pp. 639--645, 2000.

\bibitem{newman1977errors}
M.~Newman, ``An analysis of sixth-grade pupils’ errors on written mathematics,'' \emph{Research in mathematics education in Australia}, 1977.

\bibitem{clements1980errors}
M.~A. Clements, ``Analyzing children’s errors on written mathematical tasks,'' \emph{Educational Studies in Mathematics}, vol.~11, no.~1, pp. 1--21, 1980.

\bibitem{baker2008bkt}
R.~S. Baker, A.~T. Corbett, and V.~Aleven, ``More accurate student modeling through contextual estimation of slip and guess probabilities in bayesian knowledge tracing,'' in \emph{Intelligent Tutoring Systems}.\hskip 1em plus 0.5em minus 0.4em\relax Springer, Berlin, Heidelberg, 2008, pp. 406--415.

\bibitem{sanpedro2011carelessness}
M.~O. C. Z.~S. Pedro, M.~M.~T. Rodrigo, and R.~S. J.~D. Baker, ``The relationship between carelessness and affect in a cognitive tutor,'' in \emph{Affective Computing and Intelligent Interaction}, S.~D’Mello, A.~Graesser, B.~Schuller, and J.-C. Martin, Eds.\hskip 1em plus 0.5em minus 0.4em\relax Springer Berlin Heidelberg, 2011, pp. 306--315.

\bibitem{almeda2020stem}
M.~V. Almeda and R.~S. Baker, ``Predicting student participation in stem careers: The role of affect and engagement during middle school,'' \emph{Journal of Educational Data Mining}, vol.~12, no.~2, pp. 33--47, 2020.

\bibitem{sanpedro2014carelessness}
M.~O. Z.~S. Pedro, R.~S. J.~D. Baker, and M.~M.~T. Rodrigo, ``Carelessness and affect in an intelligent tutoring system for mathematics,'' \emph{International Journal of Artificial Intelligence in Education}, vol.~24, no.~2, pp. 189--210, 2014.

\bibitem{fancsali2015carelessness}
S.~E. Fancsali, ``Confounding carelessness? exploring causal relationships between carelessness, affect, behavior, and learning in cognitive tutor algebra,'' in \emph{Proceedings of International Conference on Educational Data Mining}, 2015, pp. 508--511.

\bibitem{corbett1995knowledge_tracing}
A.~T. Corbett and J.~R. Anderson, ``Knowledge tracing: Modeling the acquisition of procedural knowledge,'' \emph{User Modelling and User-Adapted Interaction}, vol.~4, no.~4, pp. 253--278, 1995.

\bibitem{gong2010comparing}
Y.~Gong, J.~E. Beck, and N.~T. Heffernan, ``Comparing knowledge tracing and performance factor analysis by using multiple model fitting procedures,'' in \emph{Intelligent Tutoring Systems}.\hskip 1em plus 0.5em minus 0.4em\relax Springer Berlin Heidelberg, 2010, pp. 35--44.

\bibitem{pardos2008composition}
Z.~Pardos, J.~E. Beck, C.~Ruiz, and N.~Heffernan, ``The composition effect: Conjunctive or compensatory? an analysis of multi-skill math questions in its,'' in \emph{Proceedings of the 1st International Conference on Educational Data Mining}, 2008, pp. 147--156.

\bibitem{pavlik2009performance}
P.~Pavlik, H.~Cen, and K.~Koedinger, ``Performance factors analysis – a new alternative to knowledge tracing,'' in \emph{Proceedings of the 14th International Conference on Artificial Intelligence in Education}, 2009, pp. 531--538.

\bibitem{baker2010gaming}
R.~S. Baker, A.~Mitrovi{\'c}, and M.~Mathews, ``Detecting gaming the system in constraint-based tutors,'' in \emph{Proceedings of the 18th Annual Conference on User Modeling, Adaptation, and Personalization}, 2010, pp. 267--278.

\bibitem{richey2021gaming}
J.~E. Richey, J.~Zhang, R.~Das, J.~M. Andres-Bray, R.~Scruggs, M.~Mogessie, R.~S. Baker, and B.~M. McLaren, ``Gaming and confrustion explain learning advantages for a math digital learning game,'' \emph{Artificial Intelligence in Education}, pp. 342--355, 2021.

\bibitem{mclaren2017game_based_learning}
B.~M. McLaren, D.~M. Adams, R.~E. Mayer, and J.~Forlizzi, ``A computer-based game that promotes mathematics learning more than a conventional approach,'' \emph{International Journal of Game-Based Learning}, vol.~7, no.~1, pp. 36--56, 2017.

\bibitem{chi1994self_explanations}
M.~T.~H. Chi, N.~D. Leeuw, M.~Chiu, and C.~Lavancher, ``Eliciting self‐explanations improves understanding,'' \emph{Cognitive Science}, vol.~18, no.~3, pp. 439--477, 1994.

\bibitem{johnson2010self_explanation}
C.~I. Johnson and R.~E. Mayer, ``Applying the self-explanation principle to multimedia learning in a computer-based game-like environment,'' \emph{Computers in Human Behavior}, vol.~26, no.~6, pp. 1246--1252, 2010.

\bibitem{wylie2014self_explanation}
R.~Wylie and M.~T. Chi, ``The self-explanation principle in multimedia learning,'' in \emph{The Cambridge Handbook of Multimedia Learning}.\hskip 1em plus 0.5em minus 0.4em\relax Cambridge University Press, 2014, pp. 413--432.

\bibitem{baker2024gender}
R.~S. Baker, J.~E. Richey, J.~Zhang, S.~Karumbaiah, J.~M. Andres-Bray, H.~A. Nguyen, J.~M. A.~L. Andres, and B.~M. McLaren, ``Gaming the system mediates the relationship between gender and learning outcomes in a digital learning game,'' \emph{Instructional Science}, August 2024.

\bibitem{mclaren2022focused}
B.~M. McLaren, J.~E. Richey, H.~A. Nguyen, and M.~Mogessie, ``Focused self-explanations lead to the best learning outcomes in a digital learning game,'' in \emph{Proceedings of the 16th International Conference on Learning Science}, 2022, pp. 36--56.

\bibitem{zambrano2024algorithmic}
A.~F. Zambrano, J.~Zhang, and R.~S. Baker, ``Investigating algorithmic bias on bayesian knowledge tracing and carelessness detectors,'' in \emph{The 14th Learning Analytics and Knowledge Conference}, 2024.

\bibitem{baker2010contextual}
R.~S. Baker, A.~T. Corbett, S.~M. Gowda, A.~Z. Wagner, B.~A. MacLaren, L.~R. Kauffman, A.~P. Mitchell, and S.~Giguere, ``Contextual slip and prediction of student performance after use of an intelligent tutor,'' in \emph{User Modeling, Adaptation, and Personalization}.\hskip 1em plus 0.5em minus 0.4em\relax Springer Berlin Heidelberg, 2010, pp. 52--63.

\bibitem{pedregosa2011scikit}
F.~Pedregosa, G.~Varoquaux, A.~Gramfort, V.~Michel, B.~Thirion, O.~Grisel, M.~Blondel, P.~Prettenhofer, R.~Weiss, V.~Dubourg, J.~Vanderplas, A.~Passos, and D.~Cournapeau, ``Scikit-learn: Machine learning in python,'' \emph{Journal of Machine Learning Research}, vol.~12, pp. 2825--2830, 2011.

\end{thebibliography}

\end{document}